\documentclass[aps,reprint,prb,amsmath,amssymb,citeautoscript,
  numerical,twocolumn,showpacs,superscriptaddress]{revtex4-1}
\usepackage{graphicx}
\usepackage{amsmath}
\usepackage{bm}

\def\bnabla{\mbox{\boldmath $\nabla$}}
\newcommand{\p}{\partial}

\begin{document}


\title{Transient stimulated emission from multi-split-gated graphene 
  structure}
\author{A. Satou}\email{Corresponding author: a-satou@riec.tohoku.ac.jp}
\author{F. T. Vasko}\altaffiliation{Present address: QK Applications,
  San Francisco, CA 94033, USA}
\author{T. Otsuji}
\affiliation{Research Institute of Electrical Communication,
  Tohoku University, Sendai 980-8577, Japan}
\author{V. V. Mitin}
\affiliation{Department of Electrical Engineering, University at 
  Buffalo, Buffalo, NY 1460-1920, USA}
\date{\today}

\begin{abstract}
  Mechanism of transient population inversion in graphene with 
  multi-splitted (interdigitated) top-gate and grounded back 
  gate is suggested and examined for the mid-infrared (mid-IR)
  spectral region. 
  Efficient stimulated emission after fast lateral spreading of 
  carriers due to drift-diffusion processes is found for the 
  case of a slow electron-hole recombination in the passive region.
  We show that with the large gate-to-graphene distance the drift 
  process always precedes the diffusion process, due to the 
  ineffective screening of the inplane electric field by the gates.
  Conditions for lasing with a gain above 100 cm$^{-1}$ are 
  found for cases of single- and multi-layer graphene placed in the 
  waveguide formed by the top and back gates. Both the waveguide
  losses and temperature effects are analyzed.
\end{abstract}
\maketitle

\section{Introduction}

Conventional scheme of semiconductor laser \cite{Koechner-Silfvast} 
is based on 
population inversion between electron states in conduction and valence
bands, so that the emission wavelength is determined by the bandgap 
of the material used (typically, lasing takes place in a spectral 
region from near-IR to far-UV). Lasers for mid-IR and THz regions 
were realized based on the tunnel-coupled heterostructures 
(quantum cascade scheme\cite{Gmachl-RPP-2001}) or on the $p$-type bulk 
materials with the degenerate valence bands~\cite{Komiyama-AP-1982}. 
Active studies 
of graphene, which is a two-dimensional gapless semiconductor with 
unusual physical characteristics\cite{CastroNeto-RMP-2009}, 
involve both theoretical investigation of the stimulated emission 
regime under steady-state or ultrafast 
pumping~\cite{Ryzhii-JAP-2007,Vasko-PRB-2010} and experimental 
attempts - approaches for realization of lasing. Recently, the 
transformation of the ultrafast optical pumping into THz or near-IR 
radiation was reported, see Refs.~\onlinecite{Boubanga-PRB-2012,
  Prechtel-NC-2012, Li-PRL-2012}. Due to
the emission of optical phonons after the ultrafast 
pumping~\cite{Boubanga-PRB-2012}, these approaches should lose an 
efficiency with 
increasing pulse duration. In order to avoid the suppression of 
stimulated emission, one needs a pumping scheme which permits to 
create dense electron-hole plasma without involvement of the 
high-energy states when the optical-phonon emission becomes essential. 
Thus, investigation of alternative pumping mechanisms for 
realization of population inversion in electron-hole plasma of 
graphene is timely now.
\begin{figure}[t]
  \begin{center}
    \includegraphics[width=8.5cm]{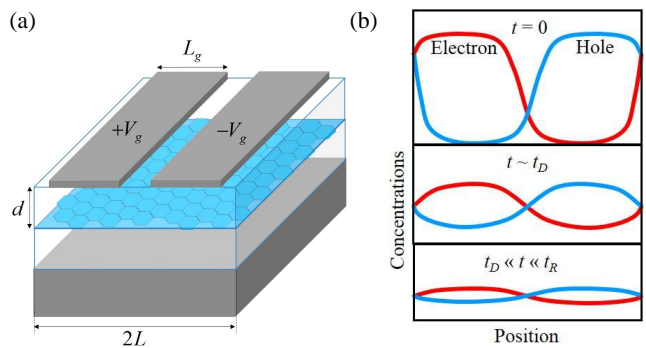}
  \caption{\label{FigSchematics}
    (Color online) (a) Multi-split-gated structure under
    initial biases $+V_{g}$ and $-V_{g}$ with a grounded back
    gate at the bottom. (b) Transient evolution of
    charge concentrations after switching off the voltages 
    $\pm V_{g}$ at $t=0$ when lateral drift and diffusion processes 
    take place during time intervals $t\sim t_D$ and 
    $t_{D} \ll t \ll t_{R}$, where $t_{D}$ and $t_{R}$ are the
    characteristic time scales of the drift-diffusion and 
    recombination processes, respectively.}
  \end{center}
\end{figure}

In this paper, we suggest a new pumping scheme for a graphene layer
modulated by spatio-temporally varied voltages applied 
through multi-splitted gates (MSG), see the structure in 
Fig.~\ref{FigSchematics}(a). An initial periodical modulation of 
electron and hole concentrations shown in 
Fig.~\ref{FigSchematics}(b) takes place at $t\leq 0$ under bipolar 
voltages $\pm V_{g}$ applied through the top gates of 
the MSG structure. 
Temporal evolution of the initial charge distributions due to 
lateral spreading of carriers after abrupt switch-off voltages 
$\pm V_{g}$ at $t\geq 0$ is shown in the middle panel of 
Fig.~\ref{FigSchematics}(b). 
If recombination processes are negligible, in-plane spreading of 
electrons and holes drifted by the electric field created 
by themselves
takes place with a time scale $t_{D}$, followed by their
diffusion within time $t_{D} \ll t \ll t_{R}$, where
$t_{R}$ is the characteristic time for non-radiative recombination,
without changing the total electron 
and hole concentrations which are determined by the initial 
conditions at $t\leq 0$.
The MSG structure also works as a waveguide where the multi-splitted
top and back gates define the vertical confinement and waves 
propagate along the waveguide.

Under a typical disorder level, the drift-diffusive hydrodynamic
equations describe regime of spreading at $t>0$, when
the initial distributions transform into homogeneous quasi-Fermi 
distributions of electrons and holes.
If transient stimulated emission due to direct interband 
transition in the spectral region $\hbar\omega\leq 2\varepsilon_F$ 
takes place during time scales less or comparable to 
the characteristic time $t_{D}$ in the 
passive region (at energies less than a half of the optical-phonon 
energy), an effective regime for the transient stimulated emission 
is accomplished.

The paper is organized as follows. In the next section we calculate 
the distributions of carriers in the biased structure under 
consideration. In Sec. III, we analyze the process of lateral 
drift and diffusion of carriers after the bias voltages are 
abruptly switched-off.
The transient lasing regime is considered in Sec. IV. The last 
section includes the list of approximations used and conclusions.
In Appendix we evaluate the hydrodynamic equations describing
a temporal evolution of non-uniform electron-hole plasma.

\section{Initial electron and hole distributions}

We start from consideration of the initial distribution of 
electron and hole concentrations under the biases $\pm V_g$ applied 
through the multi-splitted top gates separated by the distance 
$d$ from graphene placed at $z=0$ over the substrate with the
grounded bottom gate at $z=-d$. The two-dimensional Poisson equation
\begin{equation}\label{EqPoisson}
  \frac{\p^2 \varphi}{\p x^2}+\frac{\p^2 \varphi}{\p z^2}
  = 0, ~~ |z| < d, ~~ |x| <L
\end{equation}
should be supplied by periodic boundary conditions along 
the structure ($x$-direction, here $2L$ is the length of the 
two-strip element with $+V_g$ and $-V_g$ voltages), boundary 
conditions at the top gates and bottom gate,
$\varphi|_{|x\pm L/2|<L_{g}/2,z=d}=\pm V_{g}$ and $\varphi|_{z=-d}=0$,
and boundary conditions far from the structure, 
$(\p\varphi/\p z)_{x,z\to\infty} = 0$.
Boundary conditions at the graphene layer are given by the continuity 
requirement $\varphi|_{z=-0}^{z=+0}=0$ and the Gauss theorem: 
\begin{equation}\label{EqGaussTheorem}
  \left.\frac{\p \varphi}{\p z}\right|_{z=-0}^{z=+0}
  = -\frac{4\pi}{\epsilon}\rho.
\end{equation}
Here $\epsilon$ is the static dielectric constant, which is the same 
for the layers under and above the graphene layer, and 
$\rho=\rho_{e}+\rho_{h}$ is 
the total charge density in the graphene layer, where
$\rho_{e}$ and $\rho_{h}$ are the electron and hole charge densities,
respectively.

Initial distributions of the potential and the charge 
densities can be found by solving Eq.~(\ref{EqPoisson}) 
self-consistently with the following relation between the charge 
densities and the potential in the graphene layer, $\varphi|_{z=0}$:
\begin{equation}\label{EqChargeDensity}
  \rho_{r} = s_{r}\frac{4e}{2\pi\hbar^{2}}\int_{0}^{\infty}dp
  pf_{F}(p, -s_{r}e\varphi|_{z=0}, T),
\end{equation}
where $s_{e} = -1$ and $s_{h} = +1$, $T$ is the carrier 
temperature, and 
$f_{F}(p, E, T) = \{1+\exp[(vp-E)/T]\}^{-1}$ 
is the quasi-Fermi distribution with the carrier velocity
$v = 10^{8}$ cm/s.

The charge densities and potential would be obtained
from Eqs.~(\ref{EqPoisson}) and (\ref{EqChargeDensity}) 
analytically as 
$|\rho_{r}| = \rho_{s} = \epsilon V_{g}/4\pi d = C_{s}V_{g}$
and 
$\varphi = \varphi_{s} = \sqrt{\pi\hbar^{2}v^{2}C_{s}V_{g}/2e^{3}}$,
in case when the parallel-plate model could be
applicable, the temperature would be zero, and the quantum 
capacitance of graphene could be ignored.
However, the parallel-plate model is not applicable in our case,
due to the limitation of the vertical dimension $d$ that determines
the operating frequency of the waveguide structure and
the magnitude of gate voltages in order to have population inversion
up to that frequency.
Intending waveguide structures operating at mid-IR wavelengths,
in the discussion below we shall set the structural parameters as
$d=1.5-4$ $\mu$m, $L=1.5-6$ $\mu$m, $L_{g}=0.5-2$ $\mu$m,
and $\epsilon=4$ (SiO$_{2}$), and 
with thickness of the gates fixed to $10$ nm.
However, it should be mentioned that SiO$_{2}$ has sharp absorption 
peaks at $9$ and $21$ $\mu$m~\cite{Kitamura-AO-2007}. Therefore, we
shall focus on $\lambda = 12$, $15$, and $30$ $\mu$m
(corresponding to $f = 25$, $20$, and $10$ THz, respectively).
In general, either adapting non-polar waveguide materials
or avoiding absorption peaks in polar materials 
is preferable to avoid the dielectric loss in the THz/mid-IR region.

\begin{figure}[t]
  \begin{center}
    \includegraphics[width=8.5cm]{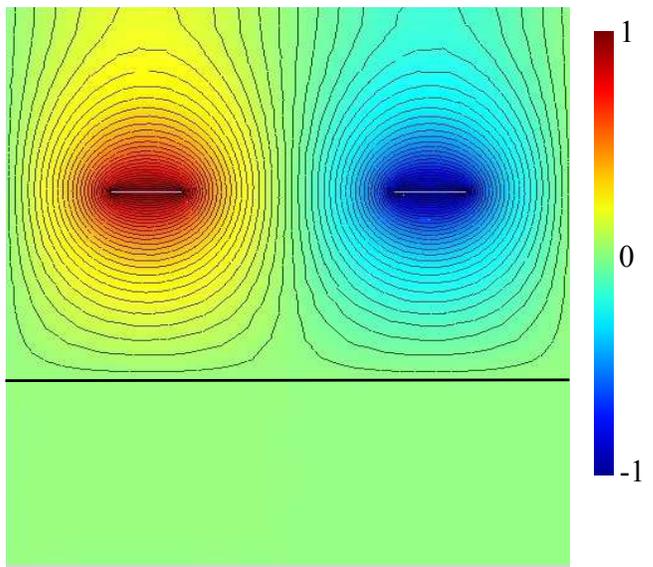}
    \caption{\label{FigInitPotential}
      (Color online) Contour plot for dimensionless potential 
      distribution, $\varphi/V_{g}$, in the structure with
      $d = 1.5$ $\mu$m, $L=2$ $\mu$m, $L_{g} = 0.5$ $\mu$m, and 
      $\epsilon=4$ 
      at $T=300$ K and $V_{g}=200$ V. The bold black line corresponds
      to the graphene layer, whereas the gray lines to the gates.
      In between the graphene layer and the bottom gate 
      there is only an invisible variation of the potential,
      of the order of $0.1$ V.}
  \end{center}
\end{figure}
\begin{figure}[t]
  \begin{center}
    \includegraphics[width=8cm]{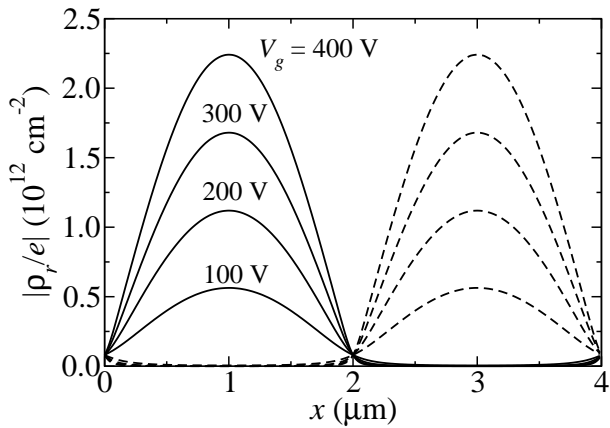}
    \caption{\label{FigInitConc}
      Distributions of electron ($r=e$) and hole ($r=h$)
      concentrations (solid and dashed lines, respectively),
      with the same parameters as in Fig.~\ref{FigInitPotential}
      except gate voltages.}
  \end{center}
\end{figure}

Figure~\ref{FigInitPotential} shows 
the normalized potential distribution, $\varphi/V_{g}$,
with $d = 1.5$ $\mu$m, $L=2$ $\mu$m, and $L_{g} = 0.5$ $\mu$m
 at $T=300$ K and $V_{g}=200$ V, and Fig.~\ref{FigInitConc} shows
electron and hole concentrations, $|\rho_{r}/e|$ 
with different gate voltages.
As expected, the gate voltages induce
electrons and holes in the graphene layer, in which the potential
is slightly fluctuated from zero to have nonzero charge densities.
Since we have the gates placed far away from the graphene layer,
the electron and hole charge densities are smaller than those
obtained using the parallel-plate model, $\rho = \rho_{s}$, 
and they have sinusoidal-like shapes. Moreover,
since we have the ratio of $L/d$ close to unity,
the electric field in the graphene layer created by one gate is 
canceled due to 
the fringe effect by the other gate with opposite-signed voltage,
so that the charge densities are further reduced.
The electric field is mostly concentrated at the edges of the gates, 
reflecting their thickness much smaller than their length.
The maximum field for the highest voltage is close to but still 
below the breakdown field of SiO$_2$, $\sim 10$ 
MV/cm.~\cite{Sze}
On the other hand, the field is almost zero in between the graphene
layer and the bottom gate due to the screening.

\section{Transient Lateral Drift-Diffusion and Population 
  Inversion}

After the abrupt switch-off of the gate voltages, the carriers 
spread over the graphene layer by the lateral drift-diffusion 
processes. In the limit of effective intercarrier scattering
and momentum relaxation due to the elastic scattering on
structural disorders, these processes are governed by the 
hydrodynamic equations described in Appendix, 
Eqs.~(\ref{EqContinuity}) and (\ref{EqCurrentDensity}),
coupled with the self-consistent Poisson equation, 
Eq.~(\ref{EqPoisson}). In this section, they were solved numerically 
as nonlinear
equations for the quasi-Fermi levels for electrons and holes,
$\mu_{r{\bf x}t}$, using the standard finite-difference scheme.
As mentioned above, we also assume that recombination processes
are suppressed.

\begin{figure}[t]
  \begin{center}
    \includegraphics[width=8cm]{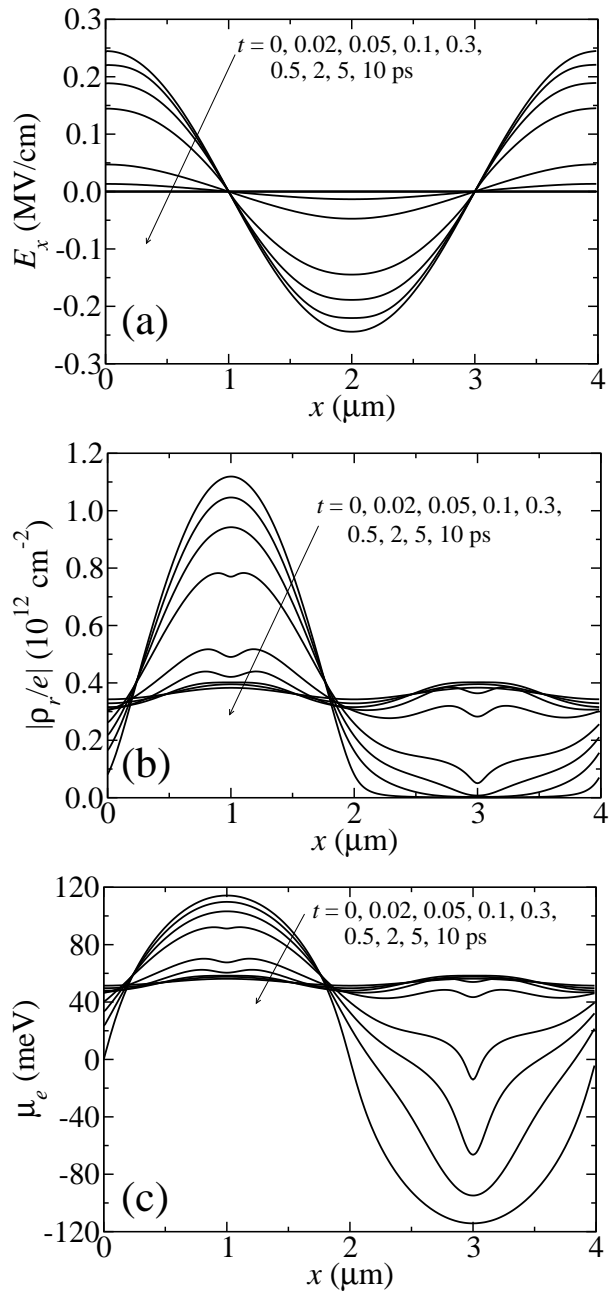}
    \caption{\label{FigTimeDep}
      Transient evolution of (a) in-plane electric field,
      (b) electron concentration, and (c) electron quasi-Fermi level
      with the same parameters as in Fig.~\ref{FigInitPotential}
      and with $\upsilon_{\text{tot}}/v = 0.5$.}
  \end{center}
\end{figure}

\begin{figure}[t]
  \begin{center}
    \includegraphics[width=8cm]{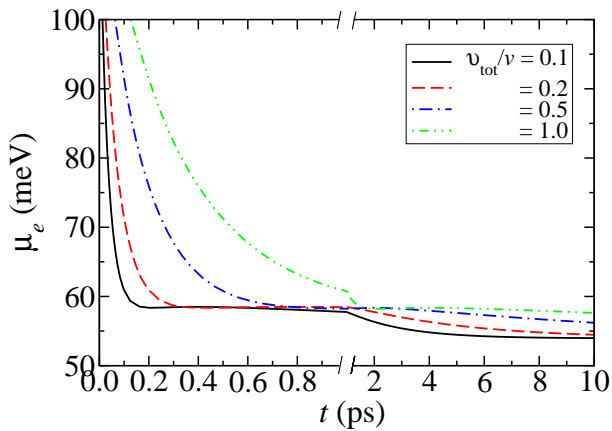}
    \caption{\label{FigMuTDep}
      (Color online) Time dependence of electron quasi-Fermi level 
      at $x=L/2$ with the same parameters as in 
      Fig.~\ref{FigInitPotential} and with different values
      of $\upsilon_{\text{tot}}/v$.}
  \end{center}
\end{figure}

Figures~\ref{FigTimeDep}(a), (b), and (c) show the time and position 
dependences of the electric field, electron concentration, and 
electron quasi-Fermi level, respectively (see also 
Fig.~\ref{FigMuTDep}). Here the same parameters
as in Fig.~\ref{FigInitPotential} were used
and, we have an additional parameter, $\upsilon_{\text{tot}}/v$,
which characterizes the total scattering rate caused by 
structural disorders (see Appendix for details).
Hole concentration and its
quasi-Fermi level are just equal to mirror images of
Figs.~\ref{FigTimeDep}(b) and (c).
They show two distinct time scales of relaxation
of these quantities towards their steady states.
One is about $0.5$ ps, and it is associated with a process that
continues until the electric field created
by carriers becomes
negligibly small, thus identified as the drift process. The other 
is about $10$ ps and is due to the diffusion process.
In our case where $d$ is comparable to $L$ and therefore
the screening of the inplane electric field by the gates is
ineffective, the time scale of the drift
process is always much faster than that of the diffusion process,
unless the quasi-Fermi level is not too large.
It can also be seen from Fig.~\ref{FigTimeDep}(c) that
more or less uniform distribution of the electron quasi-Fermi level
around $50$ meV is reached and population inversion in the THz 
range through the entire graphene layer is established just after the 
drift 
process. Qualitatively speaking, this is shorter than the time scale 
of nonradiative recombination via optical phonons, which is
in between $1-10$ ps 
(see, for example, Ref.~\onlinecite{Satou-JAP-2013}).

Figure~\ref{FigMuTDep} shows the time dependence of the
electron quasi-Fermi level at the center of the graphene layer
with different values of $\upsilon_{\text{tot}}/v$. The time scales
increase simultaneously when $\upsilon_{\text{tot}}/v$ increases,
since the carrier spreading becomes slower when its scattering
becomes more frequent. A dimensional analysis
shows that the time scales are proportional to 
$\upsilon_{\text{tot}}/v$. Moreover, they are almost proportional to
$L^{2}$ when $d$, $L_{g}$, and $V_{g}$ are scaled linearly to $L$.
On the other hand, when fixing $d$ and $V_{g}$ while scaling
$L$ and $L_{g}$, the time scale of the drift process
is roughly proportional to $L$ whereas that of the diffusion process
remains proportional to $L^{2}$.

\begin{figure}[t]
  \begin{center}
    \includegraphics[width=8.5cm]{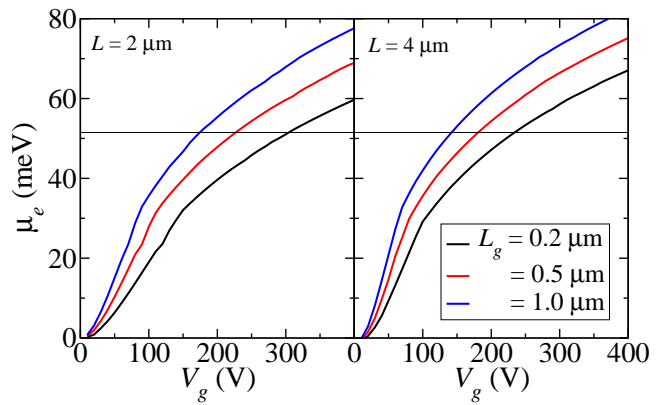}
    \caption{\label{FigMuVgLgLDep} (Color online)
      Minimum electron quasi-Fermi level at $t=0.3$ ps as a function 
      of gate voltage $V_{g}$ with different half-period lengths
      $L$ and different gate lengths $L_{g}$, together with
      $\upsilon_{\text{tot}}/v = 0.1$. Thin solid lines indicate
      a half of the photon energy corresponding to $\lambda = 12$
      $\mu$m.}
  \end{center}
\end{figure}

Figure~\ref{FigMuVgLgLDep} shows the minimum electron
quasi-Fermi level at $t=0.3$ ps as a function of gate voltage $V_{g}$ 
with different half-period lengths $L$ and different gate lengths 
$L_{g}$, together with $\upsilon_{\text{tot}}/v = 0.1$. 
As can be seen in Figs.~\ref{FigTimeDep}(c) and \ref{FigMuTDep}, 
we have more or less uniform quasi-Fermi level 
at the time $t=0.3$ ps for $\upsilon_{\text{tot}}/v = 0.1$,
so that it makes sense to discuss about a single quasi-Fermi level
(we took the minimum quasi-Fermi level to ensure that population 
inversion at a certain
energy takes place all over the graphene layer).
It is clear from Fig.~\ref{FigMuVgLgLDep} that
the quasi-Fermi level
increases monotonically as the voltage increases.
Also, it increases as the gate length increases.
The condition of population inversion, 
$\hbar\omega/2 < \mu_{r}$, for $\lambda = 12$ $\mu$m
is fulfilled at voltage $V_{g} > 150-300$ V depending on the gate 
length $L_{g}$ and the half-period length $L$.
Since we have neglected the nonradiative recombination, the 
quasi-Fermi level is solely determined by the initial concentration 
and, thus, by the electrostatics through the gate voltages and 
geometrical parameters in a non-trivial way. In particular, 
threshold voltages for the same value of the quasi-Fermi level
are lower for $L = 4$ $\mu$m than those for $L = 2$ $\mu$m.
This lowering is associated with the larger ratio of $L/d$
in the former case, where the cancellation of the electric field
in the graphene layer discussed in the previous section
is largely relaxed
and thereby the induced charge densities as well as the total charges
become larger.

From the aspect of the breakdown field, detailed numerical
analysis showed that the maximum field at the gate edges to obtain 
the same value of quasi-Fermi level (say, $60$ meV) varies a little 
by the value of $L_{g}$ for a fixed value of $L$,
although there exists an optimal value of $L_{g}$; 
for example, for $L = 2$ $\mu$m, 
the maximum field ranges from $2$ to $2.4$ MV/cm when
$L_{g} = 0.2-1$ $\mu$m, having the minimum value
at $L_{g} \simeq 0.4$ $\mu$m.
In conjunction with the fringing effect mentioned above,
the maximum field becomes several times smaller as $L$ changes 
from $L = 2$ to $4$ $\mu$m (reduced to around $0.7$ MV/cm).
Note that for too long $L$, the time scale of the drift process
becomes longer, so that the nonradiative recombination effectively
takes place before the drift process completes, and the condition of 
population inversion might not be fulfilled.
It is worth mentioning that the results obtained in this 
section depend only weakly on the temperature through the
temperature dependence of the quasi-Fermi level for fixed
charge densities.
In particular, the quasi-Fermi level becomes slightly larger as the 
temperature becomes lower. However, this is not the case for
gain of the waveguide since values of distribution functions
around the quasi-Fermi level mainly determine the gain and 
depend much on the temperature.

\section{Transient Gain of MSG Structures}

Finally, we turn to estimate gain in the MSG structure.
We determine the electric field of TE mode $E_z\exp (ikx-i\omega t)$ 
propagating along the waveguide from the wave equation 
\begin{equation}\label{EqWave}
  \left[\frac{d^2}{dz^2}-k^2
    +\left({\frac{\sqrt{\varepsilon}\omega}{c}}
    \right)^2\right]E_z = 0. \\
\end{equation}
Here we take into account the dielectric loss by introducing
the complex dielectric constant, 
$\varepsilon = \varepsilon'+i\varepsilon''$,
while we assume $\varepsilon' \gg \varepsilon''$.
We use the boundary conditions at multi-splitted and bottom gates 
$E_{z=\pm d}=0$. We set to $E_{z=d}=0$ 
the boundary condition at 
the multi-splitted gates because the gate period is shorter than
the wavelength.

At the graphene layer we use a continuity requirement 
$E_z |_{z=-0}^{z=+0}=0$ and a boundary condition
\begin{equation}\label{EqWaveBoundaryCond}
  \left. \frac{dE_z}{dz} \right|_{-0}^0 =
  -i\frac{4\pi\omega}{c^2}\sigma_\omega E_{z=0},
\end{equation}
which is written through the high-frequency conductivity of graphene, 
$\sigma_\omega =\sigma '_\omega +i\sigma ''_\omega$. 
Using the 
solution of Eqs.~(\ref{EqWave}) and (\ref{EqWaveBoundaryCond}), 
$E_z\propto\sin\kappa (d-|z|)$ with 
$\kappa =\sqrt{\varepsilon (\omega /c)^2 -k^2 }$,
one obtains the dispersion relation between the wavenumber $k$ and 
the frequency $\omega$:
\begin{equation}\label{EqDispersionRelation}
  \kappa d\cot (\kappa d) = 2\pi 
  (-\sigma ''_\omega+i\sigma '_\omega)\omega d/c^2.
\end{equation}
Since the Drude-like conductivity is negligibly small at the designed
frequencies of the waveguide, the conductivity can 
be written as
\begin{equation}\label{EqConductivity}
  \sigma '_\omega = \frac{e^{2}}{4\hbar}
  \left(1-\widetilde{f}_{e{\bf x}p=p_{\omega}t}
  -\widetilde{f}_{h{\bf x}p=p_{\omega}t}\right),
  ~~
  \sigma ''_{\omega} \simeq 0,
\end{equation}
where $p_{\omega} = \hbar\omega/2v$ and $\widetilde{f}_{r}$ is
the quasi-Fermi distributions of electrons and holes with
temperature $T$ and quasi-Fermi levels $\mu_{r}$.
Here we take the minima of the quasi-Fermi levels 
to avoid the complication by their position dependence 
in the transient behavior discussed in
the previous section.
In case of multi-layer graphene, the conductivity
roughly becomes $m$-fold larger, where $m$ is the number of
graphene layers.

Taking into account the smallness of 
$\sigma '_{\omega}/c$, we derive the explicit expression of 
the complex wavenumber from Eq.~(\ref{EqDispersionRelation}):
\begin{equation}\label{EqComplexWavenumber}
  kd \simeq \sqrt{(\sqrt{\varepsilon'}\omega d/c)^{2}
    -(\pi/2)^{2}+i\frac{\omega d}{c^{2}}
    (4\pi\sigma'_{\omega}+\varepsilon''\omega d)}.
\end{equation}
Away from the condition $\sqrt{\varepsilon'}\omega d/c = \pi/2$,
the following approximate expressions for the gain
defined as $g_{\omega}=-\text{Im}\ k$ and the real wavenumber
$k_{\omega} = \text{Re}\ k$ can be obtained from
Eq.~(\ref{EqComplexWavenumber}):
\begin{equation}\label{EqGainAndWaveNumber}
  \begin{array}{l}
    \displaystyle
    g_\omega d \simeq -\frac{\omega}{2k_{\omega}c^{2}} 
    (4\pi\sigma'_{\omega}+\varepsilon''\omega d),
    \vspace{0.1cm}
    \\
    \displaystyle
    k_\omega d \simeq \sqrt{(\sqrt{\varepsilon'}\omega d/c)^{2}
    -(\pi/2)^{2}},
    \\
  \end{array}
\end{equation}
for $\sqrt{\varepsilon'}\omega d/c > \pi/2$. This case
corresponds to a propagating mode with relatively small gain.
In the opposite case, we have a quasi-standing-wave mode
with high gain, which is not of our interest. 
Considering the nonnegligible absorption by
SiO$_{2}$, it turns out that the waveguide under consideration
has a rather narrow bandwidth for the propagating mode around the 
central frequency $\omega = \pi c/2\sqrt{\varepsilon'}d$ when the
thickness $d$ is fixed.
Writing the imaginary part of the dielectric constant $\varepsilon''$
through the absorption index $n''$ 
($\varepsilon''\simeq 2\sqrt{\varepsilon'}n''$),
the condition of positive gain can be readily obtained from 
Eq.~(\ref{EqGainAndWaveNumber}) for multi-layer graphene:
\begin{equation}\label{EqPositiveGainCond}
  -m\left(1-\widetilde{f}_{e{\bf x}p=p_{\omega}t}
  -\widetilde{f}_{h{\bf x}p=p_{\omega}t}\right)
  \gtrsim \frac{n''}{\alpha}
  \simeq \frac{\varepsilon''}{2\sqrt{\varepsilon'}\alpha},
\end{equation}
where $\alpha = e^{2}/\hbar c$ is the fine structure constant.
Since  the left-hand side in Eq.~(\ref{EqPositiveGainCond}) is 
smaller than $m$, 
we have a rather universal expression for the maximum allowed
value of the dielectric loss for positive gain, $n'' < m\alpha$.

\begin{figure}[t]
  \begin{center}
    \includegraphics[width=8cm]{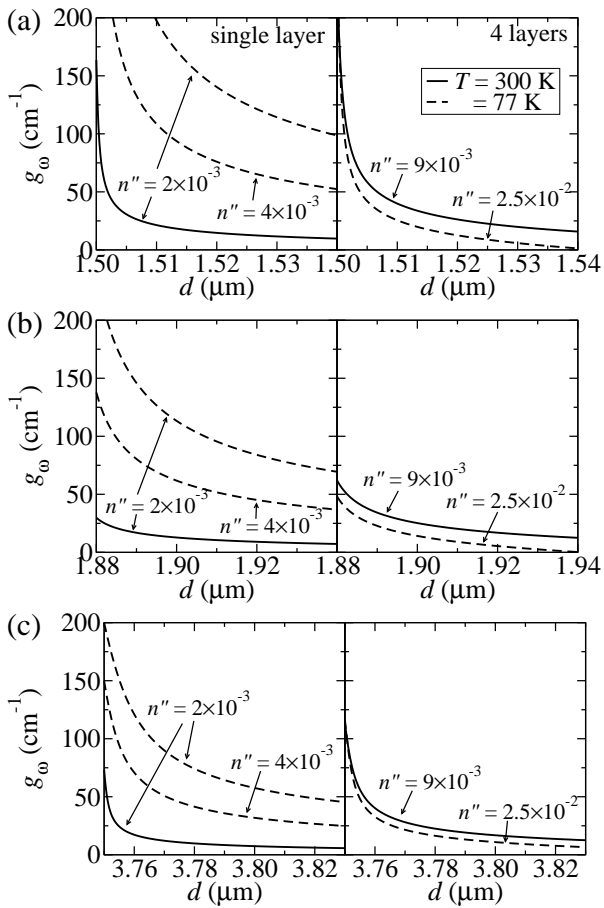}
    \caption{\label{FigGainDDep}
      Dependence of gain on the thickness $d$ 
      with operating wavelengths $\lambda = 12$, $15$, and $30$
      $\mu$m, quasi-Fermi levels $\mu = 70$, $60$, and $40$ meV
      for panels (a), (b), and (c), 
      respectively, different absorption indices,
      different temperatures, and different numbers of graphene
      layers (left panels for single layer and right for 4 layers).}
  \end{center}
\end{figure}

Figure~\ref{FigGainDDep} shows the dependence of gain 
on the thickness $d$ with operating wavelengths 
$\lambda = 12$, $15$, and $30$ $\mu$m, quasi-Fermi 
levels $\mu = 70$, $60$, and $40$ meV, respectively,
different absorption indices,
different temperatures, and different numbers of graphene layers,
 using Eq.~(\ref{EqComplexWavenumber}).
Note that for each wavelength under consideration 
the condition of population inversion is satisfied with corresponding
quasi-Fermi level.
We plotted Fig.~(\ref{FigGainDDep}) for the thickness larger than
$\pi c/2\sqrt{\varepsilon'}\omega$ ($d_{0}=1.5$, $1.88$, and 
$3.75$ $\mu$m for $\lambda = 12$, $15$, and $30$ $\mu$m,
respectively), which correspond to the propagating mode.
At the condition $d = \pi c/2\sqrt{\varepsilon'}\omega$
the values of the gain and real wavenumber coincide.
It is seen from Fig.~(\ref{FigGainDDep}) that
the gain above $100-200$ cm$^{-2}$ is achieved and that
the gain decreases as the thickness increases; 
the real wavenumber increases as understood from 
Eq.~(\ref{EqGainAndWaveNumber}).
Thus, one needs to choose carefully a proper value of the thickness 
$d$ (slightly larger than $d_{0}$) to have sufficiently large real 
wavenumber while keeping the gain larger than losses.
Conversely, once the thickness is determined, the operating 
wavelength is limited in a rather narrow-band range.

\begin{figure}[t]
  \begin{center}
    \includegraphics[width=8cm]{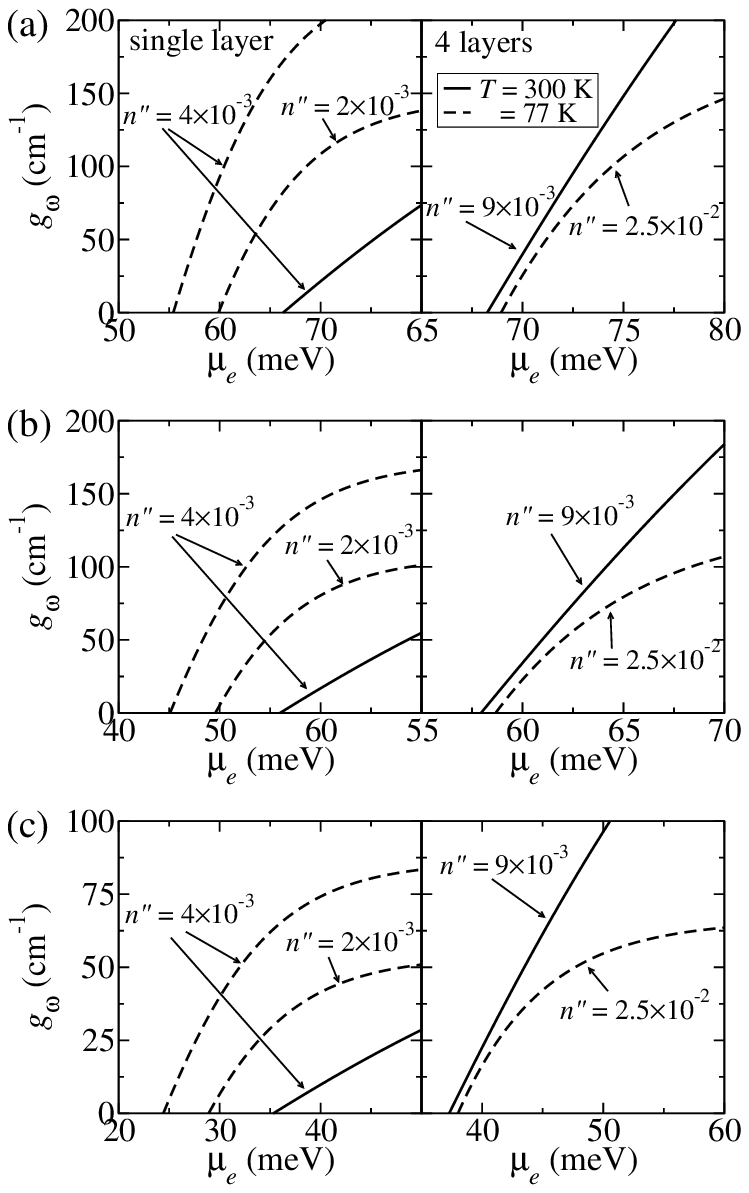}
    \caption{\label{FigGainMuDep}
      Dependence of gain on the quasi-Fermi level with fixed 
      thickness $d$ ($\lambda = 12$, $15$, $30$ $\mu$m and
      $d = 1.51$, $1.89$, $3.78$ $\mu$m for panels (a), (b), (c),
      respectively).
      Other parameters are the same as in Fig.~\ref{FigGainDDep}.}
  \end{center}
\end{figure}

Figure~\ref{FigGainMuDep} shows the dependence of gain on the 
quasi-Fermi level with a fixed thickness for each operating 
wavelength. The real wavenumber does not
noticeably change by the quasi-Fermi level.
It is seen in Fig.~\ref{FigGainMuDep} that the gain increases 
almost linearly to the quasi-Fermi level, starting from the onset of 
positive gain described by Eq.~(\ref{EqPositiveGainCond}),
and it saturates after the increase in the quasi-Fermi
level by temperature.
Owning to the factor $1/d$ in the expression of $k$, the
gain increases as the operating wavelength decreases.


Figures~\ref{FigGainDDep} and \ref{FigGainMuDep} exhibits a 
relatively strong temperature-dependence of the gain.
This reflects the fact that the interband negative
conductivity is linearly dependent on values of distribution 
functions as seen in 
Eq.~(\ref{EqConductivity}), resulting in the increase in 
its absolute value for the frequency below the quasi-Fermi level 
as the temperature decreases. Thus, at low temperature
the positive gain appears for the higher absorption index.
It in turn means that for a fixed absorption index
the threshold quasi-Fermi level for the positive gain becomes
lower at lower temperature, although above the threshold
the positive gain quickly saturates and the magnitude remains
more or less the same below $T=77$ K.
Also, the introduction of multi-layer graphene greatly enhances
the gain. The absorption index $n''$ experimentally
measured for fused silica glasses in 
Ref.~\onlinecite{Kitamura-AO-2007}
is above $10^{-2}$, although a smaller value is expected by a 
thermally grown crystalline SiO$_{2}$ on a Si substrate.
As seen in right panels of Fig.~\ref{FigGainDDep},
the dielectric loss in such a case can be overcome by introducing
multi-layer graphene and by operating at low temperature.
One can alternatively use nonpolar materials with no large 
absorption in the mid-IR wavelength as a part of
the MSG structure, e.g., in the substrate region, where
materials with relatively low breakdown field are permissible.

\section{Conclusions}

With the main goal to find conditions for an effective stimulated 
emission regime without the optical-phonon emission, we have 
examined a new pumping scheme for a graphene layer modulated by
spatio-temporally varied voltages which are applied through the 
multi-splitted top gates. We found that a transient lasing regime in 
the mid-IR spectral region is realized in the MSG structure
with micrometer width and period. Gain above hundred(s)
cm$^{-1}$ for operating wavelengths $\lambda = 12$, $15$, and 
$30$ $\mu$m was obtained if the gate voltage is around 100-300 V. 
This is 
comparable to gain in typical quantum cascade lasers
\cite{Jovanovic-APL-2004} and the transient stimulated emission 
takes place if waveguide losses are comparable to losses in quantum
cascade lasers.

Let us discuss the assumptions we used in the presented 
calculations. Because a luck of data on graphene structures with 
multi-splitted gates, the above consideration was separated into 
description of the different stages of evolution (i.e., diffusion 
from the separated electron and hole distributions at initial moment 
to the homogeneous electron-hole plasma) and estimates of the gain at
a time when the recombination is negligible (see numerical data
for recombination rate in Ref.~\onlinecite{Rana-PRB-2009}). We 
considered the simplified periodical geometry (the edge effects 
require a special consideration) placed into a media with homogeneous 
dielectric constant. A more complicate spatio-temporal simulation
does not change the numerical estimates presented here and it should 
be performed for a specific structure. 
In calculating the waveguide mode of the MSG structures we
assumed the zero field at the plane where the multi-splitted gate
is placed. Other assumptions are rather 
standard. We used the semi-phenomenological model of elastic 
scattering
under description of the diffusion process in 
Appendix, assumed the rate of long-range disorder scattering
is proportional to carrier momentum (see discussion and 
conditions in Ref.~\onlinecite{Vasko-PRB-2007}), 
and restricted ourselves by the single-particle approach.
In addition, we do not consider a nonlinear regime of lasing, so
that a pulse duration is not determined here. An estimate
of the recovery time requires a special consideration.

To conclude, we believe that the results obtained open a way for 
a further experimental investigation of the transient stimulated 
emission in the mid-IR spectral region. Note that attempts for 
realization of the graphene-based laser in the THz and near-IR 
spectral region were performed during last 
years~\cite{Ryzhii-JAP-2007, Vasko-PRB-2010, Boubanga-PRB-2012,
  Prechtel-NC-2012, Li-PRL-2012}. Similar investigations in the 
mid-IR spectral region should be useful for realization of
devices in that region.

\begin{acknowledgments}
This work was supported by JSPS Grant-in-Aid for Specially Promoted
Research (\#23000008), by JSPS Grant-in-Aid for Young Scientists
(B) (\#23760300), and by NSF TERANO award 0968405.
Also, FTV and VVM are grateful to RIEC as the major amount of this 
work was done during their visit to Tohoku University on invitation 
of RIEC.
\end{acknowledgments}

\appendix*
\section{Hydrodynamic Approach}

The electron and hole contributions ($r=e$ and $h$) to the charge 
and current densities, $\rho_{r{\bf x}t}$ and ${\bf I}_{r{\bf x}t}$, 
are determined by the standard quasi-classical relations:
\begin{equation}\label{EqChargeAndCurrent}
  \left|\begin{array}{*{20}c}
  \rho_{r{\bf x}t}  \\ {\bf I}_{r{\bf x}t} \end{array} \right| 
  = 4s_{r}e\int\frac{d{\bf p}}
  {(2\pi\hbar )^2} \left|\begin{array}{*{20}c}
  1 \\ {\bf v}_{\bf p} \end{array} \right| f_{r{\bf xp}t} ,
\end{equation}
where $f_{r{\bf xp}t}$ stands for electron or hole distributions and 
$s_{e} = -1$ and $s_{h} = +1$.
The spatio-temporal evolution of the quantities 
(\ref{EqChargeAndCurrent}) is governed by the quasi-classical 
kinetic equation for $f_{r{\bf xp}t}$, see Ref.~\onlinecite{Vasko}.
In this paper, we consider the case of effective momentum relaxation 
due to the elastic scattering on structural disorders, under the 
conditions $\nu_m\equiv
\nu_{\bar p}\gg |e{\bf E}|/{\bar p}$ (here $\bar p$ is a 
characteristic momentum) 
and $\nu_m\gg v/L$, when the distribution is given by 
$f_{r{\bf xp}t}=f_{r{\bf x}pt}+\Delta f_{r{\bf xp}t}$ with a weak 
anisotropic part, $\Delta 
f_{r{\bf xp}t}=-\Delta f_{r{\bf x,-p}t}$. The linearized equation 
for $\Delta f$ gives the anisotropic distribution in the form:
\begin{equation}\label{EqAnisoDistr}
  \Delta f_{r{\bf xp}t}= -\left({\bf v}_{\bf p}\cdot\bnabla_{\bf x}
  +s_{r} e{\bf E}_{{\bf x}t}
  \cdot\bnabla_{\bf p}\right) f_{r{\bf x}pt} /\nu_p , 
\end{equation}
where we used the elastic collision integral 
$-\nu_p\Delta f_{r{\bf xp}t}$ written through the momentum 
relaxation frequency $\nu_p$.
The isotropic part of distribution $f_{r{\bf x}pt}$ is governed by 
the equation:
\begin{eqnarray}\label{EqIsoDistr}
  \frac{\partial f_{r{\bf x}pt}}{\partial t}+\Phi_{r{\bf x}pt}=
  \sum\limits_{k} J_{k}(f_{{\bf x}t} |rp) ,  \\
  \Phi_{r{\bf x}pt} =\overline{\left({\bf v}_{\bf p}\cdot
    \bnabla_{\bf x}+s_{r} e{\bf E}_{{\bf x}t}\cdot\bnabla_{\bf p}
    \right)\Delta f_{r{\bf x}pt}} , ~~~  \nonumber
\end{eqnarray}
where overline means the averaging over ${\bf p}$-plane angle and 
summation over $k$ includes the non-elastic scattering mechanisms 
($k=ac,opt,cc$ for relaxation via acoustic and optical phonons or 
carrier-carrier scattering).

A general solution for Eqs.~(\ref{EqAnisoDistr}) and 
(\ref{EqIsoDistr}) is determined both by external field and by 
relative contributions of the nonelastic scattering mechanisms.
Below we consider the case an effective intercarrier scattering, 
under the condition 
\begin{equation}\label{EqEffectiveInterCarrier}
  \nu_m\gg\nu_{cc}\gg\nu_{ac}, ~\nu_{opt}, 
\end{equation}
where $\nu_k$ means the scattering rate for $k$th channel. 
The solution is given by the quasiequilibrium distribution:
\begin{equation}\label{EqQuasiEquilibriumDistr}
  \widetilde{f}_{r{\bf x}pt}=
  \left\{\exp [(vp-\mu_{r{\bf x}t})/T_{{\bf x}t}]+1\right\}^{-1} 
\end{equation}
written without a negligible correction of the order of 
$(\nu_{ac}, ~\nu_{opt})/\nu_{cc}$. Note that 
$J_{cc}(\widetilde{f}_{{\bf x}t} |rp)=0$ and the main term of 
Eq.~(\ref{EqIsoDistr}) vanishes by the solution 
(\ref{EqQuasiEquilibriumDistr}) 
with any effective quasi-Fermi levels, $\mu_{r{\bf x}t}$, and an 
arbitrary effective temperature, $T_{{\bf x}t}$.

In order to obtain $\mu_{r{\bf x}t}$ and $T_{{\bf x}t}$, we take 
into account that
\begin{equation}\label{EqConservation}
  \frac{4}{L^2}\sum\limits_{\bf p}\left| \begin{array}{*{20}c} 1
    \\  v p  \\
  \end{array}\right| J_{cc}\left( f_{{\bf x}t}|r{\bf p}\right) =0, 
\end{equation}
i.e., the intercarrier scattering does not change the electron and 
hole concentrations and the energy of carriers, as it follows 
from the explicit expression for $J_{cc}$.
Thus, the functions $\mu_{r{\bf x}t}$ and $T_{{\bf x}t}$ are 
determined from the balance equations for the charge and energy 
densities, while ${\bf I}_{r{\bf x}t}$ is
determined through $\mu_{r{\bf x}t}$ and $T_{{\bf x}t}$ according to 
Eqs. (\ref{EqChargeAndCurrent}), (\ref{EqAnisoDistr}), and 
(\ref{EqQuasiEquilibriumDistr}). Integrating Eq.~(\ref{EqIsoDistr}) 
over ${\bf p}$-plane, one obtains the balance equations 
for electron and hole charge densities:
\begin{equation}\label{EqContinuity}
  \frac{\partial \rho_{r{\bf x}t}}{\partial t}
  +\bnabla_{\bf x}\cdot{\bf I}_{r{\bf x}t}=
  \left(\frac{\partial \rho_{r{\bf x}t}}{\partial t}\right)_{rec},
\end{equation}
where we take into account that $4e\int d{\bf p}\Phi_{r{\bf x}pt}
/(2\pi\hbar )^2=s_{r}\bnabla_{\bf x}\cdot {\bf I}_{r{\bf x}t}$ and
the right-hand side describes the recombination 
processes~\cite{note1}.

We consider the momentum relaxation caused by Gaussian and 
short-range disorder potentials, \cite{Vasko-PRB-2007} with the 
total rate 
$\nu_p$. The Gaussian disorder is described by the correlation 
function $\overline{V}^{2}\exp \left[ -(\mathbf{x}-\mathbf{x}^{\prime
  })^{2}/2l_{c}^{2}\right]$, where $\overline{V}$ is the averaged
energy and $l_{c}$ is the correlation length. Within the Born 
approximation, the correspondent relaxation rate reads 
$\nu_p =(\upsilon _{d}p/\hbar)\Psi (pl_{c}/\hbar )(1+v_0 /v_d)$ where 
we have introduced the dimensionless function
$\Psi (z)=e^{-z^{2}}I_{1}(z^{2})/z^{2}$ with the first-order Bessel 
function of an imaginary argument, $I_{1}(z)$ and the characteristic 
velocity $\upsilon_{d}=\pi (\overline{V}l_{c}/\hbar)^{2}/(2\upsilon)$.
The relaxation rate due to the short-range disorder potential 
has a similar (if $l_{c}\rightarrow 0$) dependence $\propto\upsilon
_{0}p/\hbar$, with an explicit expression for the characteristic 
velocity $\upsilon_{0}$ given in Ref.~\onlinecite{Vasko-PRB-2007}.
Assuming that the carrier temperatures are equal to the lattice
temperature, $T_{{\bf x}t} = T$, the current density can be written
in the following expression:
\begin{equation}\label{EqCurrentDensity}
  {\bf I}_{r{\bf x}t}=
  \sigma_{r{\bf x}t}
  \left({\bf E}_{r{\bf x}t}
  -\frac{s_{r}}{e}\bnabla_{\bf x}\mu_{r{\bf x}t}\right),
\end{equation}
where the local conductivity $\sigma_{r{\bf x}t}$ is given by
$\sigma_{r{\bf x}t} \simeq (e^{2}v/\pi\hbar\upsilon_{\text{tot}})
\widetilde{f}_{r{\bf x}p=0t}$.
We assumed here that the total scattering rate is 
proportional to the momentum, i.e., 
$\nu_{p} = (\upsilon_{\text{tot}}/\hbar)p$ 
with the characteristic velocity 
$\upsilon_{\text{tot}}=\upsilon_{0}+\upsilon_{d}$
of the total scattering rate. The value of $\upsilon_{\text{tot}}$
can be estimated as $\upsilon_{\text{tot}}/v = 
6.58\times10^{-2}-10^{-1}$, which correspond to
the value of the total scattering rate $10^{13}-10^{14}$ s$^{-1}$
at $vp = 100$ meV.


\end{document}